\documentclass[twocolumn]{pasj00}
\draft

\SetRunningHead{Astronomical Society of Japan}{Usage of \texttt{pasj00.cls}}

\title{Low/Hard State Spectra of GRO J1655-40 Observed with Suzaku}

\author{Hiromitsu \textsc{Takahashi}\altaffilmark{1}, Yasushi \textsc{Fukazawa}\altaffilmark{1},
Tsunefumi \textsc{Mizuno}\altaffilmark{1}, Ayumi \textsc{Hirasawa}\altaffilmark{1} \\
Shunji \textsc{Kitamoto}\altaffilmark{2}, Keisuke \textsc{Sudoh}\altaffilmark{2}, Takayuki \textsc{Ogita}\altaffilmark{2}\\
Aya \textsc{Kubota}\altaffilmark{3}, Kazuo \textsc{Makishima}\altaffilmark{4,5}, Takeshi \textsc{Itoh}\altaffilmark{5}, \\
Arvind N. \textsc{Parmar}\altaffilmark{6}, Ken
\textsc{Ebisawa}\altaffilmark{7}, Sachindra
\textsc{Naik}\altaffilmark{7}, 
Tadayasu \textsc{Dotani}\altaffilmark{7},
Motohide \textsc{Kokubun}\altaffilmark{7}\\
Kousuke \textsc{Ohnuki}\altaffilmark{7}, Tadayuki \textsc{Takahashi}\altaffilmark{7}, Tahir
\textsc{Yaqoob}\altaffilmark{8}, Lorella \textsc{Angelini}\altaffilmark{9}\\
Yoshihiro \textsc{Ueda}\altaffilmark{10}, Kazutaka
\textsc{Yamaoka}\altaffilmark{11}, Taro
\textsc{Kotani}\altaffilmark{12}, Nobuyuki
\textsc{Kawai}\altaffilmark{12}\\
Masaaki \textsc{Namiki}\altaffilmark{13}, 
Takayoshi \textsc{Kohmura}\altaffilmark{14}, Hitoshi \textsc{Negoro}\altaffilmark{15}\\
}

\altaffiltext{1}{
Department of Physical Science, School of Science, Hiroshima University\\
1-3-1 Kagamiyama, Higashi-Hiroshima, Hiroshima 739-8526, Japan}
\altaffiltext{2}{
   Department of Physics, Rikkyo University\\
   3-34-1, Nishi-Ikebukuro, Toshima--ku, Tokyo 171--8501, Japan}
\altaffiltext{3}{
   Department of Electronic Information Systems, Shibaura Institute of Technology\\
   307 Fukasaku, Minuma-ku, Saitama-shi, Saitama 337-8570, Japan}
\altaffiltext{4}{
   Cosmic Radiation Laboratory, Institute of Physical and Chemical 
   Research (RIKEN)\\
   2-1 Hirosawa, Wako-shi, Saitama, 351-0198, Japan}
\altaffiltext{5}{
   Department of Physics, University of Tokyo\\
   7-3-1, Hongo, Bunkyo-ku, Tokyo, 113-0033, Japan}
\altaffiltext{6}{
   Astrophysics Mission Division, Research and Scientific Support
   Department of ESA, ESTEC\\
   Postbus 299, NL-2200 AG Noordwijk, The Netherlands}
\altaffiltext{7}{
   Institute of Space and Astronautical Science, JAXA\\
   3--1--1, Yoshinodai, Sagamihara, Kanagawa,  229-8510, Japan}
\altaffiltext{8}{ 
   Department of Physics and Astronomy, John Hopkins University\\
   3400 N, Charles St., Baltimore, MD 21218, USA}
\altaffiltext{9}{
   Explosion of the Universe Division, NASA Goddard Space Flight Center\\
   Greenbelt, MD 20771, USA}
\altaffiltext{10}{
   Department of Astronomy, Kyoto University\\
    Kitashirakawa-Oiwake-cho, Sakyo-ku, Kyoto, 606-8502, Japan}
\altaffiltext{11}{
   Department of Physics \& Mathematics, Aoyama Gakuin University\\
   5-10-1 Sagamihara, Kanagawa, 229-8558, Japan}
\altaffiltext{12}{
   Department of Physics, Tokyo Institute of Technology\\
   2-12-1, O--okayama,Meguro-ku, Tokyo, 152-8551, Japan}
\altaffiltext{13}{
   Department of Earth and Space Science, Graduate School of Science,
   Osaka University\\
   1-1, Machikaneyama, Toyonaka, Osaka, 560-0043, Japan}
\altaffiltext{14}{
   Physics Department, Kogakuin University\\
   2665--1, Nakano--cho, Hachioji, Tokyo, 192--0015, Japan}
\altaffiltext{15}{
   Department of Physics, College of Science and Technology, Nihon
   University\\
   1-8, Kanda-Surugadai, Chiyoda-ku, Tokyo, 101-8308, Japan}

\email{hirotaka@hirax7.hepl.hiroshima-u.ac.jp}


\Received{$\langle$reception date$\rangle$}
\Accepted{$\langle$acception date$\rangle$}
\Published{$\langle$publication date$\rangle$}
\begin{document}
\maketitle

\begin{abstract}
The Galactic black-hole binary GRO J1655$-$40, 
a source harboring superluminal jets, 
was observed with Suzaku on 2005 September 22--23,  
for a total time span of $\sim 1$ day, 
and a net exposure of 
35 ks with the  X-ray Imaging Spectrometer (XIS) and 
20 ks with the Hard X-ray Detector (HXD). 
The source was detected over a broad and continuous energy range of  0.7--300 keV,
with an intensity of $\sim$50~mCrab at 20 keV.
At a distance of 3.2 kpc,
the  0.7--300 keV luminosity is $ \sim 5.1 \times 10^{36}$  erg s$^{-1}$
($\sim 0.7$ \% of the Eddington luminosity for a 6 $M_{\odot}$ black hole).
The source was in a typical low/hard state,
exhibiting a power-law  shaped continuum 
with a photon index of $\sim 1.6$.
During the observation,
the source intensity gradually decreased by  25\% 
at energies above $\sim 3$ keV, and by 35\% below 2 keV. 
This, together with the soft X-ray spectra taken with the XIS,
suggests the presence of an independent soft component
that can be represented by emission from a cool ($\sim 0.2$ keV) disk.
The hard X-ray spectra obtained with the HXD reveal a 
high-energy spectral cutoff, 
with an e-folding energy of $\sim 200$ keV.
Since the spectral photon index above 10 keV is 
 harder by $\sim 0.4$ 
than that observed in the softer energy band,
and the e-folding energy is higher than those of typical reflection humps,
the entire 0.7--300 keV spectrum cannot be reproduced
by a single thermal Comptonization model,
even considering reflection effects.
Instead, the spectrum (except the soft excess)
can be successfully explained by invoking two 
thermal-Comptonization components with different $y$-parameters.
In contrast to  the high/soft state spectra of this object
in which  narrow iron absorption lines
are detected  with  equivalent widths of 60--100~eV, 
the present XIS spectra bear no such features
beyond an  upper-limit equivalent width of 25~eV. 
\end{abstract}

\clearpage

\section{Introduction}

X-ray spectra of  black-hole binaries (BHBs)
depend on the luminosity in complex ways,
particularly when they are luminous
(e.g., \cite{Fender04, Kubota04}).
However, as long as the luminosity
is well below the Eddington limit, $L_{\rm E}$,
BHBs exhibit only two major spectral states;
the high/soft state and the low/hard state,
which manifest themselves when the luminosity is
higher and lower than a few percent of $L_{\rm E}$,
respectively \citep{Tanaka96,  McClintock03}.
In the high/soft state, a BHB spectrum consists of optically-thick
thermal emission from a standard accretion disk (e.g., \cite{Makishima86}),
and a power-law shaped hard tail extending
up to MeV energies with a photon index of  ${\it \Gamma} \sim 2.3$.

In the low/hard state,
a BHB emits a  harder spectrum that can be approximated by  a
${\it \Gamma}  \sim 1.7$ power law,
accompanied by subsidiary features such as
iron-K  related structures (e.g., \cite{Ebisawa96}),
a low energy excess (e.g., \cite{B-Church95}),
and a hard X-ray hump attributable to
reflection by cool matter (e.g., \cite{Frontera01, Lightman88}).
In addition, the spectrum often exhibits a  cutoff
at energies above  50--150 keV,
as  observed from several BHBs
and a fair number of Seyfert galaxies (e.g., \cite{Quadrelli03,Risaliti02}).
As a result, a thermal-Comptonization scenario
(e.g., \cite{Sunyaev80,Poutanen96,thcomp})
and its variants have been developed as a standard interpretation.
According to this scenario,
the low/hard state emission from an accreting black hole is formed
when some seed photons are Compton up-scattered by
a hot electron distribution, with the
electron temperature determining the high-energy spectral cutoff.

In spite of its general success,
details of the thermal Comptonization scenario
are far from established.
It is yet to be clarified
whether the supposed hot electron cloud should be identified
with an optically-thin portion of the accretion disk,
or a sort of corona above an optically-thick cool disk.
The radial and vertical extents of the hot cloud are not well constrained.
Nor is it clear yet whether the soft seed photons
come from an optically-thick and cool portion of the disk,
or from  the hot cloud itself as thermal cyclotron emission.
In order to better quantify this promising scenario for the low/hard state,
careful modeling of the spectra over a broad energy band is of 
particular importance.
In fact, \citet{Frontera01} found that broad-band spectra of Cyg X-1,
obtained with  BeppoSAX,
can be successfully reproduced by invoking {\it two} Compton $y$-parameters,
of $0.15 $ and $0.89$.
We may then utilize the extremely broad energy band of Suzaku 
\citep{Mitsuda06},
to extend such studies to  a larger number of BHBs.
Here, we describe Suzaku results on the low/hard state of GRO J1655$-$40,
which  is a secure BHB having a much higher  inclination  
angle than Cyg X-1 ($20^\circ \sim 65^\circ$; \cite{Gies1986,Ninkov87}).

GRO J1655$-$40 is a transient X-ray
source discovered with the BATSE
onboard the Compton Gamma Ray Observatory \citep{Zhang94}.
Subsequent optical observations
revealed a 2.6 d binary period \citep{Bailyn95},
and successfully determined the mass
of the compact object and mass-donating star as
$5.5-7.9~M_{\odot}$ and $1.7-3.3 ~M_{\odot}$, respectively \citep{Shahbaz99}.
The system is thus considered to be a BHB.
Together with GRS 1915+105 \citep{Mirabell94},
this object is classified as an interesting subgroup of BHB
possessing superluminal radio jets \citep{Hjellming95}.
The system inclination, $i$,  of GRO J1655$-$40
is estimated to be  $85^{\circ}$ from jet kinematics,
or $\sim 70^{\circ}$ from optical light curves \citep{Hooft98}.
Although the distance is estimated to be 3.2$\pm$0.2 kpc
from the radio jet kinematics \citep{Hjellming95},
a shorter distance of $\lesssim 1.7$ kpc has been proposed by
\citet{Foellmi06} based on the spectral type of the secondary star.
In addition, GRO J1655$-$40 allowed the first detection of narrow X-ray
absorption lines from  ionized iron \citep{Ueda98, Yamaoka01}.
Together with those in other
high-inclination  sources \citep{Kotani00, Ueda01}
and in dipping binary sources \citep{DiazTrigo06},
the detection of these absorption lines indicates
the existence of highly ionized plasmas above the accretion disks.

On 2005 February 17,
an X-ray brightening of GRO J1655-40 was detected
by the RXTE All Sky Monitor~(ASM)  \citep{Markwardt05};
since then it remained active for more than half a year.
X-ray observations were carried out with
XMM-Newton \citep{DiazTrigo07, Sala07},
INTEGRAL \citep{DiazTrigo07, Shaposhnikov07},
Swift \citep{Brocksopp06},  Chandra \citep{Miller06},
and RXTE \citep{Shaposhnikov07},
as the source evolved from the low/hard state into the high/soft state,
and then back again.
Based on high energy spectra obtained in the low/hard state early in the
outburst, \citet{Shaposhnikov07} reported the detection of a
high-energy cutoff around 180 keV.
In contrast, broad-band spectra taken during the high/soft state
reveal a hard tail component extending up to $\sim 150$ keV
\citep{Brocksopp06, DiazTrigo07},
reconfirming a previous OSSE detection \citep{Kroeger96, Zhang97a}.
The absorption lines from highly ionized ions were also reconfirmed
with a blue shift equivalent to 300-6000 km s$^{-1}$
\citep{Miller06, DiazTrigo07, Sala07}.

The present paper describes Suzaku observations  of
GRO~J$1655-40$ made in the decay phase during the
2005 September campaign,
while the object was in the low/hard state.
We successfully detected the object over
a broad, unprecedented, energy range of 0.7--300 keV.
Although the obtained broad-band spectra can be approximated
by a power-law of photon index $\sim 1.6$,
the high-quality Suzaku data clearly reconfirm the  spectral
cutoff at energies above $\sim 100$ keV
as well as a weak soft X-ray excess.
Furthermore, the data require two Comptonized components, 
having a common electron temperature but different optical depths.
Absorption features around 6 keV were not detected significantly.

\section{Observation}

\begin{figure}
\begin{center}
\rotatebox{0}{
\FigureFile(80mm,){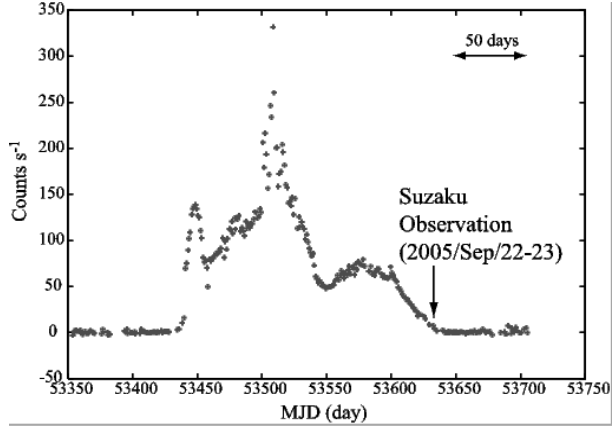}
}
\end{center}
\caption{
An X-ray (1.5--12 keV) light curve of GRO~J1655$-$40 obtained by the RXTE ASM.
The date of the Suzaku observation is indicated by an arrow.
}
\label{fig:asmlc}
\end{figure}

Using Suzaku \citep{Mitsuda06},
we observed GRO~J1655$-$40
from 2005 September 22, 07:16 UT, 
through 07:07 UT of the next day.
The data were acquired with
the X-ray Imaging Spectrometer (XIS: \cite{Koyama06}) 
and the Hard X-ray Detector (HXD: \cite{hxd1,hxd2}).
As shown in figure~\ref{fig:asmlc}, 
the source was at that time in the decay phase
of the 2005 outburst, exhibiting a typical 1.5--12 keV
intensity of a few tens mCrab.
 
Installed at the focal plane of the four X-ray telescopes (XRT:
\cite{Serlemitsos06}), 
the XIS consists of one back-illuminated CCD
(BI-CCD) camera and three front-illuminated CCD (FI-CCD) cameras,
observing in the energy range from 0.2 keV to 12 keV.  
During the present GRO~J1655$-$40 observation, 
the XIS was operated in the normal 3$\times$3 and 2$\times$2 modes.  
Since the source is relatively  bright, 
we employed ``1/8 window'' option in which, out of the 1024 rows, 128 rows 
around the source image are read out.  
This shortens the frame exposure time to 1 s, 
and reduces the number of XIS events suffering from pile-up.  
Furthermore, one of the FI-CCDs (XIS0) was not used
in order to double the telemetry capacity of the BI-CCD (XIS1).  

The HXD was operated in the normal mode, acquiring 10--70 keV data with
the Si PIN photo diodes (hereafter PIN) and 40--600 keV data with the
GSO scintillators (hereafter GSO).  Combining PIN and GSO with the XIS,
Suzaku can cover a broad energy band spanning three orders of magnitude.

The average 0.7--300 keV flux of GRO~J1655$-$40 was measured
to be $4.0 \times 10^{-9}$ erg s$^{-1}$ cm$^{-2}$.
The luminosity in the same band
was estimated to be $5.1 \times 10^{36}$ erg s$^{-1}$,
which corresponds to $\sim 0.7$ \% of $L_{\rm E}$ for a 6 $M_{\odot}$ black hole,
assuming a distance of 3.2 kpc \citep{Hjellming95}.

\section{Data Analysis and Results}

\subsection{Data Selection and Background Subtraction}
For the XIS data analysis,
we accumulated cleaned events over good time intervals
that were selected by removing spacecraft passages through
the South Atlantic Anomaly (SAA) and 
periods of low earth elevation ($<20^\circ$ for the day-time earth and 
$<5^{\circ}$ for the night earth).  
The details are described in \citet{Fujimoto06}.
The source and background events were accumulated 
inside and outside the central 
circular region of radius $4'.3$ for each CCD image, respectively. 
Since XIS2 and XIS3 have almost the same properties, 
we hereafter present
results obtained from their co-added spectra.
The total exposure time achieved was 35 ks per sensor.

Although the intensity of GRO J1655$-$40 was
$\sim$50~mCrab and we used the ``1/8 window'' option,
we still need to be careful about photon pile-up in the XIS data.
In order to quantify this effect,
we tentatively extracted another spectrum from a $1'.0-4'.3$ annular region,
where the count rate for each pixel is several times lower than 
that within the central $1'.0$, 
and the pile-up effect is considered to be negligible in the annulus.
Comparing this spectrum with the original one
from the entire $4'.3$ radius,
we found that the original spectrum has $\sim$ 3\% and 5\% higher count rates
than the annular one in the 7--8 keV and 9--10 keV energy bands, respectively,
after equalizing their normalizations in the 1--2 keV band.
This result suggests that the region inside $1'$ slightly suffers 
from photon pile-up.
Therefore, considering the statistical error ($\sim$ 3\%) 
of the spectra after binning,
we limited the XIS energy range to $<8$ keV.

Since the HXD does not have offset detectors \citep{hxd1},
subtraction of non X-ray background is 
an essential part of the data analysis.
Generally, the background spectrum to be subtracted is synthesized,
employing certain models that are based on
the HXD data 
accumulated during periods of Earth occultation  \citep{hxd2}.
The models are being developed  by the HXD team, 
and the results are being made publicly available
\footnote{http://www.astro.isas.jaxa.jp/suzaku/analysis/hxd/hxdnxb/};
some early  Suzaku publications (e.g., \cite{Reeves06,Miniutti07})
have actually used the modeled PIN backgrounds.
However, GRO J1655$-$40 was observed only 2 months after the launch,
when long-lived activation components in the GSO background
spectra were still increasing at a significant rate. 
Therefore, the model background is thought to be
less reliable in this  particular case, compared to 
other HXD observations conducted on  later occasions.
We have hence chosen another method, 
namely  to derive the background spectrum directly
from some blank-sky observations  conducted on an occasion close
in time to the present observation of  GRO J1655$-$40.
Below, our primary method is to employ this blank-sky subtraction,
but we will compare our results to
the ``standard'' modeled background as a cross confirmation.

To carry out the blank-sky subtraction,
we employed data from an observation of the planetary nebula BD+30$^\circ$3639, 
because it was conducted on the day just before
the GRO J1655$-$40 observation (2006 September 21), 
and no hard X-ray source is catalogued in this sky region. 
The on-source (i.e., GRO J1655$-$40) and blank-sky (i.e., BD+30$^\circ$3639)
data were both accumulated after removing time periods immediately 
before (3 minutes) and after (7 minutes) each passage through the SAA, 
and those periods when Suzaku was in regions of 
low geomagnetic cutoff rigidity ($<8$  GV).  
To exclude the periods of Earth occultation, 
we further screened the on-source data by the condition 
that the target elevation above the Earth's horizon should be $> 5^{\circ}$.  
Then, the background events of PIN and GSO were  accumulated 
over the same orbital phases of Suzaku as the on-source data integration.  
The obtained on-source exposure was corrected for dead time,
using so-called pseudo events implemented in the HXD \citep{hxd1};
this  method has been verified in orbit using other 
independent dead-time indicators \citep{hxd2}.
As a result, a  live time of 20 ks was attained. 

\begin{figure}
  \begin{center}
\rotatebox{0}{
\FigureFile(80mm,){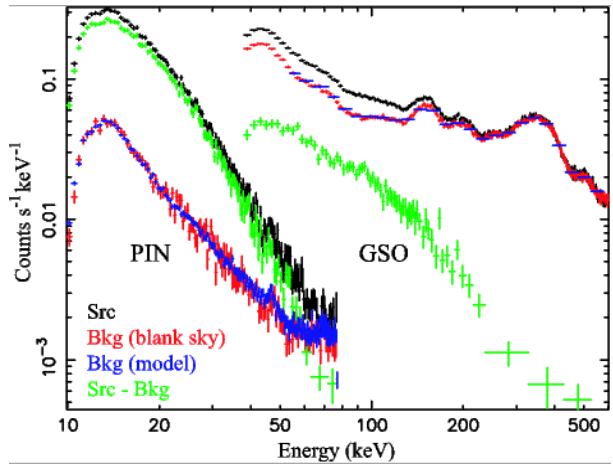}
}
  \end{center}
  \caption{
Background-subtraction procedure for the HXD data.
The on-source data, the blank-sky background data,
the modeled background data, and the background-subtracted spectra
are presented in black, red, blue, and green, respectively.
Here and hereafter,
the blank-sky data are adopted as the HXD background.
}
\label{fig:hxd_src_bkg_spec}
\end{figure}

Figure~\ref{fig:hxd_src_bkg_spec} compares
the on-source and blank-sky HXD spectra, both obtained in this way. 
The background-subtracted spectra are also presented.  
The source is so bright that the signal from GRO~J1655$-$40 
is more than half of the PIN background even at its upper-bound energy of 70 keV.
Therefore, systematic errors of the PIN background are 
considered to be negligible over the entire PIN energy range (10--70 keV).
In the GSO range, 
the source signal is 25\% and 10\% of the background 
at  40--100 keV and 100--300 keV, respectively. 
As described in the Appendix, 
the GSO background estimation using blank-sky 
observations is thought to be accurate to  about 2\%. 
As a result, we can claim a source detection
with the GSO over an energy range of 40--300 keV.  
Since the GSO background subtraction becomes more susceptible
to systematic errors as the integration time becomes shorter, 
we hereafter report results from analysis of the GSO data
integrated over the entire 20 ks exposure only.

In addition to the above analysis using  the blank-sky observation,
we tried currently available background models
to synthesize the PIN and GSO backgrounds.
These models  provide a set of  time-tagged {\it fake} events
(separately for PIN and GSO),
in the same form as the actual on-source data acquired.
We processed these background events
in exactly the same manner as the on-source data,
using in particular the same  ``good time intervals''. 
The derived model backgrounds are 
also presented in figure~\ref{fig:hxd_src_bkg_spec} .
Thus, the two sets of backgrounds from the two methods 
(the blank-sky data and the synthetic model)
agree well with each other;
the GSO backgrounds agree within  $\lesssim$ 2\%,
while the  PIN spectra agree within $\lesssim$ 3\% below $\sim$ 50 keV,
although there is a slight deviation at higher energies.

\subsection{Energy Spectra}
%
Figure~\ref{fig:broadbandspectrum} shows broad-band energy spectra of
GRO~J1655$-$40 from 0.7 keV to 300 keV, obtained with the XIS and HXD,
by  averaging all the data.
In order to assess the overall shapes of these spectra
without relying on the detector responses 
and in a way that is rather free from possible calibration uncertainties, 
we first converted them into so-called ``Crab ratios'', 
whereby the source spectrum is divided by a Crab spectrum
that was accumulated in the same way.  
As shown in figure~\ref{fig:crab_ratio}, 
the calculated Crab ratio indicates
that the object had an intensity of $\sim 50$ mCrab at  20 keV.
The Crab ratio exhibits a nearly constant logarithmic slope 
of $\sim$ 0.5 in the 2--100 keV range.  
Considering that the Crab spectrum can be approximated by a
power law with a photon index of ${\it \Gamma} \sim$ 2.1, 
this implies that the spectrum of GRO~J1655$-$40 
also takes a power-law shape, but with ${\it \Gamma} \sim$ 1.6.  
We hence conclude that GRO~J1655$-$40 was in the low/hard state
during the Suzaku observation.
In detail, the slope in the 10--70 keV band is slightly harder
than that in 3--10 keV.
The strong drop in the Crab ratio below 1 keV
is caused by a higher absorbing column density toward GRO~J1655$-$40 
than that to the Crab.

At energies above $\sim$ 100 keV, 
the Crab ratio in figure~\ref{fig:crab_ratio} exhibits a clear turnover.  
Since the Crab spectrum, 
which is about an order of magnitude brighter 
than that of GRO~J1655$-$40 at $\sim 100$ keV, 
is known to exhibit a single power-law shape in
the hard X-ray range up to 300 keV or more (e.g., \cite{Kuiper01}), 
the hint of a turnover in the Crab ratio suggests
an intrinsic high-energy cutoff in the GRO~J1655$-$40 spectrum,
as is often observed from other  BHBs in the low/hard sate. 
In the figure, the ratios are also plotted by artificially 
changing the GSO background intensity by $\pm$2\%,
which is the typical error as described in the Appendix.
Thus, the presence of a turnover itself is
unaffected by the GSO background uncertainty,
although the steepness of the cutoff might be affected.

\subsection{Light Curves}

Figure~\ref{fig:lightcurve} shows background-subtracted light curves of
GRO~J1655$-$40 obtained with the XIS and HXD-PIN.  
The intensity  decreased gradually during this one-day long observation, 
by $\sim 35\%$ in the softest 0.7--2 keV band and by $\sim 25\%$ 
in the higher energy bands, 
resulting in a slight spectral hardening with time.

In order to quantify the intensity-correlated spectral changes, 
we divided the data into two halves, which we hereafter refer to 
as the first and second halves (figure~\ref{fig:lightcurve}). 
We then extracted the corresponding energy spectra from the XIS and PIN.  
(As mentioned above, we did not attempt such
time-resolved spectral analysis on the GSO data.)  
The time coverage was half a day for each of the first and 
second halves of the PIN data, 
and the blank-sky PIN background was accumulated 
over the corresponding portions of the BD+30$^\circ$3639 observation.

Figure~\ref{fig:ratio_half} shows spectral ratios between the two halves.  
The source intensity decreased by $\sim$ 10\% 
above $\sim 3$ keV, and by up to  $\sim$ 18\% below $\sim 3$ keV.
Thus, the intensity decrease is somewhat more prominent at low energies, in
agreement with the impression from the light curves. 
 (These percentages are smaller than are indicated by the light curves, 
simply because of time averaging effects.)  
This suggests the presence of a separate soft spectral component, 
which is superposed on the dominant power-law shaped continuum
that is emitted over the entire energy band.

\subsection{Spectral Model Fitting}
\subsubsection{The XIS spectra}

As a first-cut quantitative spectral study,
we tried an absorbed single power-law model
on the XIS data of GRO~J1655$-$40,
as suggested by the Crab ratio in figure~\ref{fig:crab_ratio}.
Specifically, we fitted the  spectrum from the BI-CCD (XIS1)
and that from the two FI-CCDs (XIS2 and XIS3 co-added) simultaneously,
using the detector responses ae\_xi1\_20060213c.rmf and
ae\_xi1\_xisnom4\_20060415.arf for XIS1,
and averages of ae\_xi[23]\_20060213.rmf
and ae\_xi[23]\_xisnom4\_20060415.arf for the co-added FI spectrum.
Here and hereafter,
fits to the XIS spectra utilize only the 0.7--1.7 keV
and 1.9--8 keV ranges,
because the excluded energy bands are still subject to
calibration uncertainties or possible pile-up effects.
The effect of contamination on the optical blocking filters of
the XIS is modeled as additional absorption \citep{Koyama06}.
In the fitting, we kept the energy offset to be a free parameter
within $\pm10$~eV,
which is the current tolerance of the XIS energy scale.
The overall model normalization is here and hereafter
allowed to differ between different XIS sensors.

The single power-law fit,
with a free photon index ${\it \Gamma}$ and a free absorbing column 
density $N_{\rm H}$
(using the ``wabs'' model; \cite{Morrison83}),
was  formally rejected by the XIS data with $\chi^2/\nu = 1924/1424$,
mainly due to wiggling residuals at energies below $\sim 1.5$ keV.
This, together with  the soft-energy behavior revealed by 
figure~\ref{fig:ratio_half},
inspired us to incorporate an independent soft component.
Evidently, the most natural candidate is an optically-thick disk 
emission component
which is described by the  ``disk-blackbody (diskbb)''
model \citep{Mitsuda84, Makishima86}.
We therefore adopted a model consisting of a power-law and a disk blackbody,
both absorbed by a common absorber with column density $N_{\rm H}$.
This composite model,  denoted by ``diskbb+pow'',
gave a much improved fit with $\chi^2/\nu = 1738/1422$,
yielding an innermost disk temperature of  $kT_{\rm in} \sim 0.26$ keV
and an innermost disk radius of
$r_{\rm in} \sim 10 {/\sqrt{\cos(i)}}$ km at a distance of 3.2 kpc,
where $i$ is the inclination angle of the disk.
Nevertheless,  the model was not yet fully acceptable,
because of some positive residuals in the 5--8 keV energy range.

Finally, we obtained $\chi^2/\nu =1499/1419$,
by adding a  broad ($\sigma = 1.04^{+0.13}_{-0.10}$ keV) Gaussian 
line component,
centered at $E_{\rm center} = 6.50\pm0.08$ keV and
having an equivalent width (EW) of $280\pm40$ eV.
The fit is in fact marginally unacceptable at the 99\% confidence level,
but the fit statistic has a similar value to that obtained
when we fit the Crab spectrum with an absorbed single power-law model.
Thus, we considered that this fitting is essentially acceptable
under the current level of instrumental calibration and response adjustments.
Although we cannot rule out the possibility of
this Gaussian feature having an instrumental origin,
the obtained Gaussian parameters are suggestive
of a broad iron line, or some related spectral features.
The obtained best-fit parameters are listed in table~1.
The power-law component has  ${\it \Gamma}= 1.75$,
which is typical of an accreting black hole in the low/hard state.
The ``diskbb'' component has an innermost disk temperature of $0.18$ keV,
and carries  10.2\% of the 0.5--2 keV source flux
(or 4.1\% in the 0.5--10 keV range).

We used numerical modeling  \citep{Koyama06}
to represent the effects of contamination
on the optical blocking filters of the XIS sensors.
The model,  which is still under development,
may include some unknown systematic errors.
However, this should not affect our results significantly,
since the contaminant was not so thick
when the present observation was performed (only two months  after the launch).
Also, we did not  use data below energies of 0.7~keV,
since there is a relatively high interstellar column density toward
GRO~J1655$-$40.

\subsubsection{The HXD spectra}

Before actually conducting model fitting to
the HXD-PIN and GSO spectra of GRO~J1655$-$40,
we briefly describe the HXD data of the Crab Nebula.
With the current PIN response (ae\_hxd\_pinxinom\_20060814.rsp),
the 10--70 keV PIN spectrum of the Crab Nebula
can be successfully
represented by a single power-law model with
${\it \Gamma} = 2.09\pm0.01$,
which agrees with numerous previous measurements.
Similarly, using the GSO response  ae\_hxd\_gsoxinom\_20060321.rsp,
the 100--300 keV GSO spectrum of the Crab
is reproduced to a reasonable accuracy  with $ {\it \Gamma} \sim 2.1$
\citep{hxd2}.

Strictly speaking,
the single power-law fit to the
GSO spectrum of the Crab over the 70--300 keV range,
using the current GSO response,
is formally unacceptable,
due to mild ($\sim 10\%$) convex residuals
[see figure~18 in \citet{hxd2}].
This implies that the GSO response
predicts spectra that are slightly too 
concave than it should. 
Therefore, the GSO response needs to be improved,
or at least corrected,
when we try to quantify the high-energy cutoff of GRO J1655$-$40
suggested by its Crab ratio.
In order to make the necessary correction in an explicit manner,
we have resorted to introducing an empirical correction factor of the form
\begin{equation}
C(E) \equiv 1.36 \: (E/100)^{0.65}\exp(-E/230)~~,
\end{equation}
where $E$ is the energy in keV,
and multiplying this factor to any input model
before it is convolved with the current GSO response.
This function, $C(E)$, has a mildly  convex shape with a peak at $\sim$ 150 keV,
and  takes values in the range $0.75-0.93$  for $70<E<300$.
In fact, this  particular functional form has  been determined
so that  the current GSO response can successfully reproduce
the 70--300 keV spectrum of the Crab Nebula
by a single power-law with $ {\it \Gamma} = 2.1\pm0.1$,
when the input power-law model is multiplied by $C(E)$.
Hereafter in the present paper,
we always include this factor $C(E)$ when fitting  GSO spectra.

With these preparations,
we fitted the HXD (PIN and GSO) spectra of GRO~J1655$-$40,
after subtracting the BD+30$^\circ$3639 spectra
as the blank-sky backgrounds (subsection 3.1).
A single power-law model
(but multiplied by equation (1) when simulating the GSO data)
failed to give an acceptable joint fit to the PIN and GSO spectra,
with $\chi^2/\nu = 145/68$, even though the relative
PIN versus GSO normalization was left free to vary.
Since the model was actually over-predicting the GSO data above 100 keV,
we replaced the power-law model with a cutoff power-law
(hereafter ``cutoffpl'') model,
namely, a power-law multiplied by an exponential factor
of the form $\exp(-E/kT_{\rm cut})$,
where $k$ is the Boltzmann constant
and $T_{\rm cut}$ is the cutoff ``temperature''.
This model gave an acceptable ($\chi^2/\nu = 64/67$)
joint fit to the PIN and GSO spectra of GRO~J1655$-$40,
with the relative PIN versus GSO normalization in agreement
within  10\% of that obtained from  the Crab analysis.
The best-fit parameters are listed in table~1,
including in particular values of ${\it \Gamma} =1.35\pm0.04\pm0.05$
and $kT_{\rm cut} = 200^{+50, +140}_{-30, -60}$ keV,
where the first errors are statistical
and the second ones reflect the 2\% systematic uncertainty
in the GSO background subtraction.
The systematic errors may be somewhat overestimated,
because the value of 2\% is rather conservative (see Appendix).

By means of model fitting to the HXD data,
we have thus confirmed quantitatively
the presence of the high-energy spectral turnover
which is suggested by the Crab ratio.
Although the upper bound on $kT_{\rm cut}$ is rather loose
when considering the GSO systematics,
an $F$-test indicates  that the data still prefer
  the ``cutoffpl'' model  to the single power-law model,
because the probability of the cutoff being insignificant is 
4$\times$10$^{-7}$,
even if we adopt the worst case of 2\% under estimation of the background.
If  the correction factor descried by equation (1) is not incorporated,
the requirement of the high energy cutoff by
the HXD data for GRO~J1655$-$40 is even stronger.

With $kT_{\rm cut}\sim 200$ keV,
the  ``cutoffpl'' model as determined above from the HXD data
implies a slope of ${\it \Gamma} \sim 1.35$ in the 1--10 keV range.
This is smaller by  $\sim 0.4$
than that determined by the XIS ($\S$~3.4.1).
This reconfirms the inference
obtained by inspecting the Crab ratio (\S~3.2, figure \ref{fig:crab_ratio}),
that HXD-PIN measures a flatter slope than the XIS.

As a consistency check,
we repeated the same analysis using the HXD data,
except that the model background was subtracted ($\S$~3.1)
instead of the blank-sky data.
The obtained parameters from this analysis are
${\it \Gamma} =1.39\pm0.04$
and $kT_{\rm cut} = 290^{+100}_{-60}$ keV.
As expected from the good agreement between the two sets
of backgrounds (figure~\ref{fig:hxd_src_bkg_spec}),
neither the goodness of fit nor the best-fit parameters
differ significantly between the two methods,
reinforcing the reality of the high-energy cutoff.

\subsubsection {Broad-band spectra}

Following the above analysis,
we next jointly fitted the time-averaged XIS and HXD (PIN and GSO) spectra.
Although it would be natural to employ  a model
consisting of  ``diskbb'' and ``cutoffpl''  components,
here a more physically consistent approach may be adopted.
Since the hard X-ray emission from BHBs in the low/hard state is
generally interpreted as resulting from thermal Comptonization of
some soft photons by hot electrons with a temperature $kT_{\rm e}$ ($\S$~1),
we decided to represent the hard continuum
with ``compps'' model \citep{Poutanen96} in xspec,
instead of the ``cutoffpl'' model.
The ``compps'' code  computes Compton scattering
in an exact manner incorporating the Klein-Nishina effect,
and is known to accurately reproduce the  Wien peak
particularly when the electron temperature is high.

Our new fitting model thus becomes ``diskbb+compps+gau'',
where the broad Gaussian component around 6 keV ($\S$~3.4.1)
is retained with its parameters left free.
Because the  ``compps'' code allows us to use
an optically-thick disk emission  as a seed photon source
generated at the center of the hot plasma,
we substitute the ``diskbb'' temperature
for this seed-photon temperature.
Then, the ``compps''  normalization,
equivalent to the seed photon flux,
becomes proportional to the square of innermost radius
$r_{\rm in}^{\rm comp}$ of the assumed seed disk,
although this parameter, $r_{\rm in}^{\rm comp}$, can  differ from the original 
$r_{\rm in}$.
The implied picture is that a fraction  [$\propto (r_{\rm in})^2$]
of photons from a disk-blackbody source
is  directly observed as the soft excess,
while the rest  [$\propto (r_{\rm in}^{\rm comp})^2$]
is injected into the Comptonizing plasma.
With  the seed photon source thus specified,
the ``compps'' model has two basic free parameters:
one is $kT_{\rm e}$, and the other is so-called Compton $y$-parameter
which  is related to the optical depth $\tau$ by 
\begin{equation}
y \equiv 4 \tau \frac{kT_{\rm e}}{m_{\rm e}c^{2}}~~,
\end{equation}
where $m_{\rm e}c^{2}$ is the electron rest mass energy.
Among several available geometries provided by ``compps'',
we chose a sphere (parameter=4),
so as to make the Comptonized spectrum independent of the inclination.

We thus fitted this ``diskbb+compps+gau'' model jointly
to the  XIS, HXD-PIN, and HXD-GSO spectra,
again incorporating equation (1) for  the GSO data. 
The overall model normalization was allowed to take
different values between the XIS and HXD-PIN,
to ``absorb'' residual calibration uncertainties.
In contrast, considering the results obtained in section~3.4.2,
and the fact that the GSO scintillators are placed directly behind 
the PIN diodes,
we constrained the  PIN versus GSO normalization to stay
within $\pm$ 10\% of the nominal value determined from the Crab spectra.
This tolerance reflects  the systematic GSO background uncertainty
of $\pm 2\%$ (see Appendix),
which amounts to $\sim \pm 10\%$ of the source flux
in the  PIN/GSO overlapping energy range.
Hereafter in this section, we show only the results
using the blank-sky background,
since the modeled background yields consistent results.

\begin{figure}
\begin{center}
\rotatebox{0}{
\FigureFile(80mm,){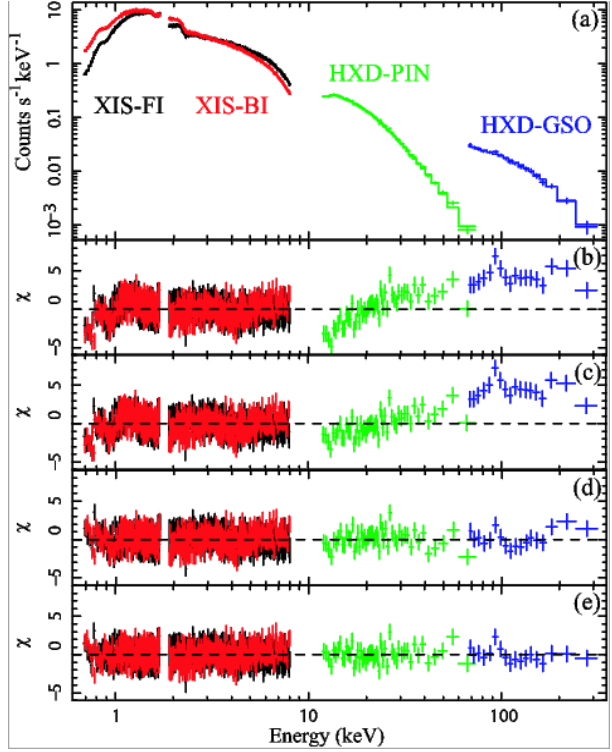}
}
\end{center}
\caption{ (a) Background-subtracted broad-band (0.7--300 keV) 
energy spectra of GRO~J1655$-$40, 
obtained with the XIS, HXD-PIN (green) and HXD-GSO (blue).  
The spectra of the FI-CCDs (black) 
and BI-CCD (red) are plotted separately.  
The fitting utilized the XIS data in the 0.7--1.7 and 1.9--8 keV ranges,
the 12--70 keV PIN data, and the GSO data in the 70--300 keV band.
The lower four panels show  residuals relative to (b) the ``diskbb+compps+gau'' model,
(c) ``diskbb+compps+reflection+gau'',
(d) ``diskbb+compps+compps+gau'',
and (e) ``diskbb+compps+compps+reflection+gau''.
 } 
\label{fig:broadbandspectrum}
\end{figure}

\begin{figure}
  \begin{center}
\rotatebox{0}{
\FigureFile(80mm,){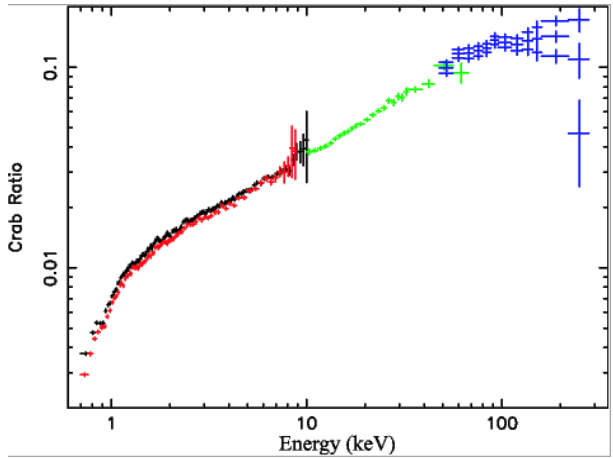}
} 
 \end{center}
  \caption{
 Time-averaged and background-subtracted spectra of  GRO~J1655$-$40,
divided by the corresponding spectra of the Crab Nebula.
GSO data after artificially changing the normalization
of the blank-sky background by
 $\pm$2\% are also shown.
The color code is the same as that in figure~\ref{fig:broadbandspectrum}.
}
\label{fig:crab_ratio}
\end{figure}

\begin{figure}
  \begin{center}
\rotatebox{0}{
\FigureFile(80mm,){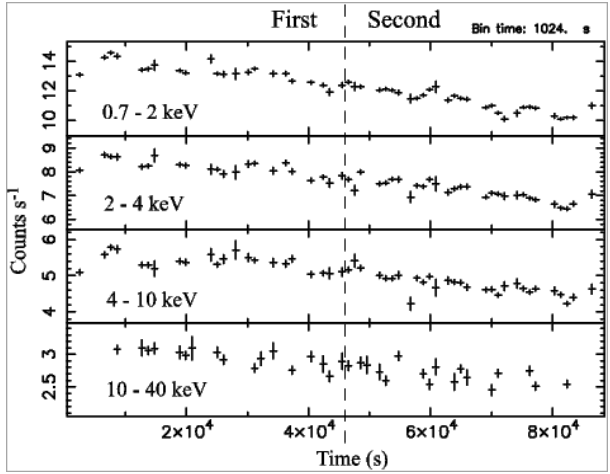}
}
  \end{center}
  \caption{
Background-subtracted light curves of GRO~J1655$-$40
obtained with the XIS1 (0.7--2, 2--4, and 4--10 keV) and PIN (10--40 keV). 
 A vertical line divides the exposure into ``first'' and
``second'' halves. 
} 
\label{fig:lightcurve}
\end{figure}

\begin{figure}
  \begin{center}
\rotatebox{0}{
\FigureFile(80mm,){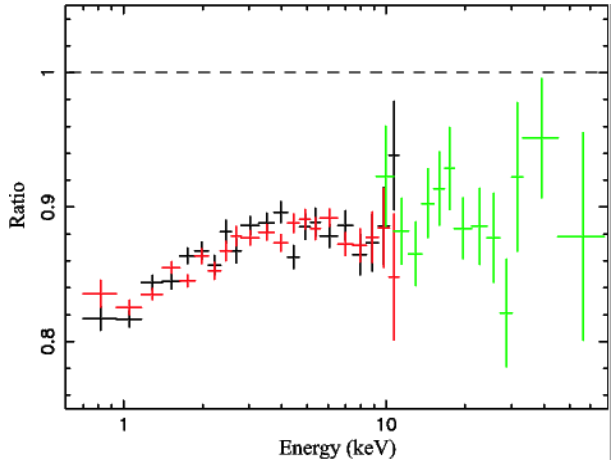}
}
  \end{center}
  \caption{ The background-subtracted spectra of GRO~J1655$-$40 obtained
 with the XIS and PIN in the second half, divided by those in the first half.}  
\label{fig:ratio_half}
\end{figure}

As presented in figure~\ref{fig:broadbandspectrum}b,
this ``diskbb+compps+gau'' model
provided a rather poor joint fit to the three spectra,
with $\chi^{2}/\nu = 2408/1489$ (table~2).
This is because the fit residuals for the PIN and GSO data are seen to rise 
toward higher energies,
yielding $\chi^{2}/\nu=576/70$
if the jointly determined model is compared with the HXD data only.
This again implies
that the continuum slope above 10 keV is harder than that in the XIS band,
as  found  previously with  the Crab ratio
and from the separate model-fitting of the XIS (${\it \Gamma} = 1.75$) and 
the HXD (${\it \Gamma} =1.35$) spectra.
Although the fit  is unacceptable,
the electron temperature obtained is $kT_{\rm e} \sim 60$ keV,
implying a cutoff energy of $3 kT_{\rm e} \sim 180$ keV 
(since the optical thickness of the ``compps'' is rather large),
which is consistent with $kT_{\rm cut}$
derived from the HXD data using the ``cutoffpl'' model.
The ``diskbb'' parameters became  different
from those derived using the XIS data only,
because we replaced the  power-law with a  ``compps''  model.

The failure of the above fit suggests the presence of a hard spectral hump,
due to reflection from thick matter
(e.g., \cite{Lightman88,Ebisawa96}).
We therefore activated an option in the ``compps'' model,
to  include reflection of the Comptonization continuum by a thin disk.
We set free three parameters specifying the reflector;
the solid angle $\Omega/2\pi$ viewed from the photon source,
its photo-ionization parameter $\xi$ in units of erg cm s$^{-1}$,
and  its  inner radius  $R_{\rm in}$ in units of $R_{\rm s}$
where $R_{\rm s}$ is Schwarzschild radius.
The reflector was assumed to have an outer radius of 1000 $R_{\rm s}$,
and a radial emissivity profile of $\propto r^{-q}$,
where $r$ is the radius and $q=3$ is  emissivity index \citep{Fabian89}.
The remaining  reflector parameters,
namely the inclination angle, abundance,  and temperature,
were fixed at $70^{\circ}$, 1 solar, and $10^6$ K, respectively.
The broad Gaussian line around 5--8~keV is included keeping the
parameters free, which is here interpreted as an iron fluorescent line 
associated with the
reflection emission.
As shown in figure~\ref{fig:broadbandspectrum}c and table~2,
this  ``diskbb+compps+reflection+gau'' model favored  $\xi \sim 0$
to represent the hard spectra in the HXD band.
However, the  fit was not improved significantly,
with $\chi^2/\nu = 2246/1486$ (0.7--300 keV) and 484/70 (10--300 keV).
The hard excess still remained above $\sim$ 50 keV
because the reflection hump cannot become hard enough
above several tens of keV.

\begin{figure}
  \begin{center}
\rotatebox{0}{
\FigureFile(80mm,){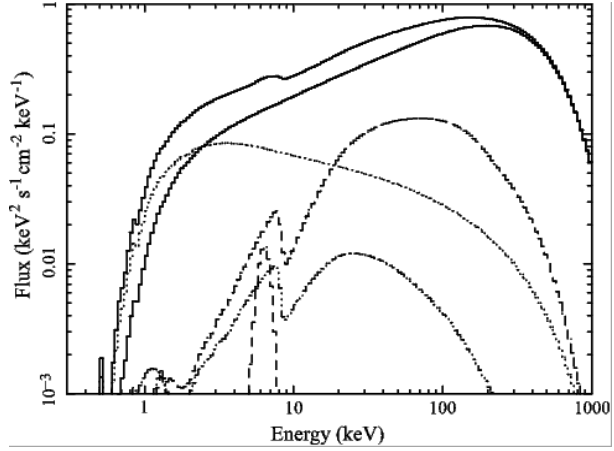}
}
  \end{center}
  \caption{
The  ``diskbb+compps+compps+reflection+gau'' models
utilized in figure~\ref{fig:broadbandspectrum} e.
Each spectral component is shown individually.
}
\label{fig:fit_model}
\end{figure}

To better reproduce the broad-band spectrum,
the model must predict a  harder and slightly concave continuum.
Therefore, as in \citet{Frontera01},
we included an additional  ``compps'' component,
instead of the reflection component.
The two ``compps'' components were assumed to share the same $kT_{\rm e}$,
but were allowed to have independent values of $y$.
As shown in figure~\ref{fig:broadbandspectrum}d and table~2,
this  ``diskbb+compps+compps+gau'' model has improved the fit to 
$\chi^2/\nu =  1600/1487$ and 85/70
in the total (0.7--300 keV) and the HXD (10--300 keV) band, respectively.
With the electron temperature of $\sim 55$ keV,
the obtained Compton $y$-parameters, $\sim 0.5$ and $\sim 1.3$,
translate to optical depths of $\sim 1.2$  and $\sim 3.0$, respectively.
The Gaussian component has an equivalent width of $\sim$110~eV.
The above two-``compps'' model requires a broad Gaussian,
which is generally considered to be an indication of a reflection
component. Therefore, we further modified the model, assuming that
the two ``compps'' components are both accompanied by reflection
with a common set of reflection parameters.
The obtained parameters are listed in table~2,
and the residuals from the best fit model are shown in figure \ref{fig:broadbandspectrum} e.  
This ``diskbb+compps+compps+reflection+gau'' model gave
$\chi^2/\nu =  1524/1484$ (0.7--300 keV) and 67/70 (10--300 keV).
Thus, the inclusion of the reflection component has significantly
improved the fit, and makes it acceptable at the 90\% confidence limit.
Also this  reflection component may 
naturally explain the  Gaussian component.  
Therefore, we  adopt this model in the following part of  this paper.
The Compton $y$-parameters, $\sim 0.28$  and $\sim 1.3$,
are almost the same as those from the ``diskbb+compps+compps+gau'' fit,
but the obtained $kT_{\rm e} \sim 140$ keV is twice as high,
thus halving the optical depths to $\sim 0.25$  and $\sim 1.2$ respectively.
As  shown in figure~\ref{fig:fit_model},
this fit implies a relatively strong reflection 
component with $\Omega/2\pi = 0.6\pm 0.1$.
This makes  $kT_{\rm e}$ higher,
because the spectral cutoff above $\sim 100$ keV
is partially explained away by the  turn-over intrinsic to the 
reflection component.

\subsubsection{Changes in the broad-band spectra}
Following the above analysis that was
based on data averaged over the whole observation time,
we applied the same
 ``diskbb+compps+compps+reflection+gau'' model separately 
to the first-half and second-half spectra,
 obtained in \S~3.3 by the XIS and HXD-PIN.
(As noted before, we did not split the GSO data into the two halves,
and the PIN data refer to those using the blank-sky background only.) 
The reflection parameters ($\Omega$, $\xi$ and $R_{\rm in}$) 
were fixed to the best-fit values of the average spectrum (table~2).
As summarized in table~3, this model 
has successfully reproduced both spectra.
As the 0.7--300 keV source luminosity  decreased by 10\%,
the two Compton $y$-parameters stayed nearly constant,
while  $kT_{\rm e}$  increased marginally.
The additional 8\% decrease (total $\sim 18\%$) fractional decrease 
of the flux below 3~keV (figure~\ref{fig:ratio_half}) is driven
solely by a decrease in the diskbb component, although the error
of the luminosity of the disk component is large.

\subsection{Narrow Iron Line Structure}

\begin{figure}
\begin{center}
\rotatebox{0}{
\FigureFile(80mm,){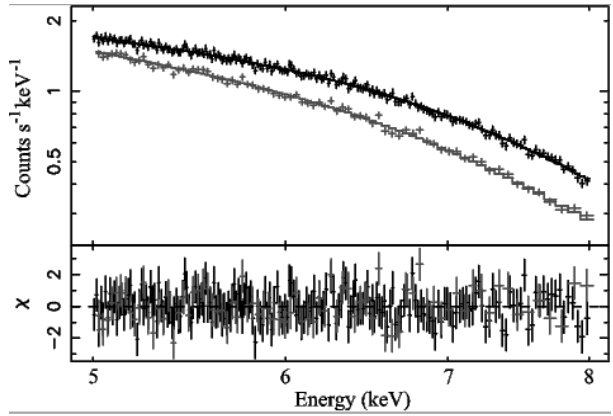}
}
\end{center}
\caption{
An expanded presentation of the
5--8 keV portion of the XIS (FI-CCD in black and BI-CCD in grey)  spectra.
The bottom panel shows the  residuals of the data 
relative to the best-fitting
model consisting of a  single power-law and  a broad
 Gaussian line.
}
\label{fig:ironlinespectrum}
\end{figure}

\begin{figure}
\begin{center}
\rotatebox{0}{
\FigureFile(80mm,){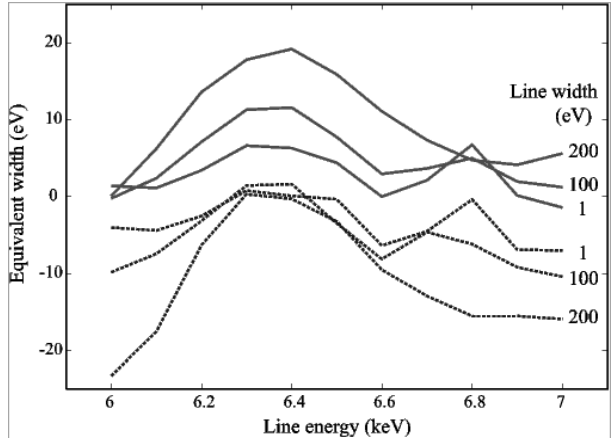}
}
\end{center}
\caption{The 90\% confidence upper limits on the equivalent widths 
of Gaussian  emission (solid) or absorption (dotted) lines in the XIS spectra,
shown as  a function of the assumed line-center energy.
}
\label{fig:ironlineeqwidth}
\end{figure}

Although the XIS data might be consistent with the presence
of a broad ($\sigma \sim 1.0$ keV) iron line feature ($\S$~3.4.1),
much narrower emission/absorption lines are often observed from BHBs.
We therefore searched the XIS spectra for relatively narrow iron K absorption or emission features.
Figure~\ref{fig:ironlinespectrum} shows an expanded view 
of the 5--8 keV portion of the XIS spectra, 
where a power-law model with a broad Gaussian mentioned in \S 3.4.1 
was fitted to the data over this limited energy range.  
As is evident from the fit residuals plotted
in the bottom panel of figure~\ref{fig:ironlinespectrum},
no narrow  spectral features are present.  
We added another Gaussian line to the model, 
allowing its normalization to have both positive and negative values.  
By fixing the centroid energy of the added Gaussian  
at various trial values between 6.0 and 7.0 keV,
we determined the allowed regions of its Gaussian equivalent width EW.  
The results are shown in figure~\ref{fig:ironlineeqwidth}, 
where three cases of the Gaussian intrinsic width were examined:
1 eV, 100 eV, and 200 eV in the Gaussian standard deviation,
based on the reports by \citet{Sala07} 
($\sim110$~eV, $\sim200$~eV, and $\sim250$~eV)
and \citet{DiazTrigo07} ($\sim 100$ eV).  
The allowed EWs of positive (emission line) and negative
(absorption line) features, at the 90\% confidence level,
are less than 20~eV and 23~eV, respectively, for 
an intrinsic width of 200~eV.
If we consider the energy range above 6.4 keV,
the upper limits of the EW are
$\sim$ 20~eV and $\sim$ 15~eV
for the emission and the absorption features, respectively.

\section{Discussion}
%
We have presented high-quality broad-band (0.7---300 keV)
energy spectra of the black hole candidate GRO J1655$-$40,
obtained with  Suzaku in a decay phase of the 2005 outburst.
Employing the distance of 3.2 kpc and assuming isotropic emission,
the 0.7--300 keV luminosity
at the time of our observation (estimated 
using the ``diskbb+compps+compps+reflection+gau'' model)
becomes $\sim 5.1 \times 10^{36}$ erg s$^{-1}$,
equivalent  to $\sim$ 0.7\% of $L_{\rm E}$ for a 6 $M_{\odot}$ black
hole.
The spectra are dominated by  a  power-law like continuum,
implying that the object was in a typical low/hard state.
The continuum is approximated by a  photon index of  $\sim 1.75$ and $\sim 1.35$ 
at energies below and above $\sim$ 10 keV, respectively,
together with a high-energy cutoff  around an energy of $\sim 200$ keV.
The spectra do not exhibit narrow iron emission or absorption lines in the 5--8 keV band,
but they do show a weak soft excess.
The continuum cannot be reproduced by a single thermal Compton model with disk reflection,
but can be explained successfully  by two thermal-Comptonization components
having a common $kT_{\rm e}$  and different optical depths, 
both accompanied by reflection with $\Omega/2\pi \sim 0.6$.
Below, we discuss these spectral characteristics.

\subsection{The High Energy Cutoff \& Two Comptonization Components}

Thanks to the high capability of Suzaku, we have shown 
that the high-energy spectrum of GRO~J1655$-$40 gradually deviates 
above $\sim 100$ keV from a single power-law,
exhibiting  a high-energy cutoff effect 
that can be represented  by an exponential factor 
 with  $kT_{\rm cut}  \sim 200 $ keV  (or $>110$~keV, if we include a
 maximum systematic error).
This reconfirms the RXTE and INTEGRAL detection of 
a similar effect with  $kT_{\rm cut} \sim 180$ keV, 
that was made in an early stage of the same outburst \citep{Shaposhnikov07}.
At that time the object was also in the low/hard state,
with a similar luminosity of $\sim 4 \times 10^{36}$ erg s$^{-1}$.
These two independent observations firmly establish the presence 
of a high-energy cutoff in the low/hard state spectrum of GRO J1655$-$40.

Although the presence of a high-energy cutoff 
supports the thermal Compton scenario,
the observed 3--80 keV continuum exhibits a  concave shape in a log-log plot,
in contradiction to a prediction of a simple Comptonization scenario
that  the continuum should have a constant logarithmic slope
at energies  $\ll kT_{\rm e}$.
The discrepancy cannot be attributed to 
reflection by cold or partially ionized material,
since it can produce a concave curvature only up to $\sim 50$ keV.
Instead, two thermal-Comptonization components  
(both modeled by ``compps''), 
with different Compton $y$-parameters,
have successfully explained the 0.7--300 keV spectra
over the entire energy band
(except for the subtle soft excess).
As shown in figure~\ref{fig:fit_model},
the flatter continuum in the HXD-PIN band is explained
by the component with the
larger Compton $y$-parameter ($y\equiv y_{\rm L}\sim 1.3$)
that has a  flatter photon index of  $\sim 1.4$,
whereas the continuum steepening by $\Delta$$\it{\Gamma}$ $\sim 0.4$ in the XIS band is 
explained by the addition of the 
component with the smaller Compton $y$-parameter ($y\equiv y_{\rm
S} \sim 0.28$)
that has 
a steeper photon index of $\sim 2.1$.
Of course, there remains  a possibility
that the two values of  $y$ are representing 
more complex, for example continuous,
distributions in the $y$-parameter.

In our double-$y$ modeling,
we assumed for simplicity that 
the two Compton components have the same $kT_{\rm e}$.
This does not  however rule out the possibility
that they have different values of $kT_{\rm e}$,
since the component with smaller $y$ contributes little 
to the GSO energy range, 
and hence its high-energy curvature remains poorly constrained. 

The empirical modeling with multiple (at least two) Comptonized
continuum components 
allow two alternative physical interpretations. 
One is a spatial variation,
in which an extended seed-photon source is 
covered by a hot electron cloud
for which the optical depth varies with position; 
the two components differ in their spatial
location but arise effectively simultaneously in time.
The other is  a temporal variation,
in which the optical depth of the hot electron cloud
varies on short timescales between, for example, two representative values,
so that a time-averaged spectrum would be composed
of a superposition of the individual components.
The true physics may well be a mixture of these two extreme cases.
For further elucidation,
we need to study intensity-correlated spectral changes on various timescales.
For example, according to \citet{Negoro01},
the 1--60 keV spectrum of Cyg X-1 acquired with Ginga softens 
around the peak of X-ray ``shots'', namely rapid X-ray flaring (see also
\cite{Poutanen99}).
This sort of investigation, however, is beyond the scope of the present paper.

\subsection{Comparison of the Compton Parameters with Cyg X-1}

The existence of two Compton components with different $y$-parameters
was previously reported from Cyg X-1 in a low/hard state observed with
BeppoSAX \citep{Frontera01},
when the 0.5--200 keV luminosity was $\sim 2.0 \times 10^{37}$ erg s$^{-1}$.
Employing the same ``compps'' modeling,
these authors measured the Compton parameters of Cyg X-1
as $y_{\rm S} \sim 0.15$ and $y_{\rm L} \sim0.89$,
together with $kT_{\rm e} \sim 60$ keV.
The implied optical depths are $ \sim 0.5$ and $\sim 1.9$.
Thus, the double-$y$ modeling has been successful on both these objects,
and has yielded roughly similar sets of spectral parameters
except for minor differences.

Compared with these values for Cyg X-1,
the two $y$-parameters of GRO~J1655$-$40 observed with Suzaku,
particularly $y_{\rm L}$, are considerably larger.
This is a direct consequence of the fact
that the BeppoSAX spectrum of Cyg X-1 is steeper
than that of GRO~J1655$-$40 obtained with Suzaku.
The difference in $y$ can be attributed to
the higher electron temperature of GRO~J1655$-$40 (140 keV),
than that of Cyg X-1 observed with BeppoSAX (60 keV).
This difference in $kT_{\rm e}$, in turn,
may arise because Cyg X-1 was by a factor of 4 
more luminous than GRO~1655$-$40.
In fact, it has been observed in other BHBs
(e.g., Esin et al. 1998; Yamaoka et al. 2005)
that $kT_{\rm e}$ decreases with the increasing luminosity,
presumably due to the enhanced coronal cooling.
This tendency, though marginal, is visible even
between the two halves of the present data of GRO~1655$-$40.
However, this mechanism alone may not be sufficient
to produce the observed temperature difference,
since the two objects are implied to have
relatively similar luminosities 
when normalized to their Eddington luminosity,
given the mass of $\sim 6~M_\odot$ of GRO~J1655$-$40
and $\sim 20~M_\odot$ of Cyg X-1 \citep{Ziolkowski05}.

In spite of the rather extreme edge-on inclination,
GRO~J1655$-$40 are inferred to have comparable,
or even smaller, Compton optical depths (1.2 and 0.25, if we adopt
$kT_{\rm e}~\sim~140$ keV for both the Comptonizing clouds.)
than Cyg X-1 (1.9 and 0.5).
This would not happen if the Comptonizing cloud
had an oblate geometry extending along the disk plane.
Therefore, the relation between the two objects is
consistent with the generally accepted view
that the Comptonizing cloud has a relatively
large scale height perpendicular to the disk plane,
so that the emergent hard X-rays are approximately isotropic.

We have so far ascribed the spectral difference 
between GRO~J1655$-$40 and Cyg X-1
to their differences either in the inclination angles or in the luminosities. 
In reality, both these effects can be operating,
that would cause GRO~J1655$-$40 to have larger $y$-parameters.
Although this subject is important in constructing 
a unified picture of BHBs in the low/hard state,
further examination of it is beyond the scope of the present paper.

\subsection{The Seed Photons and Soft Excess}
In the present analysis, we have assumed 
that the two ``compps'' (Comptonized continuum) 
components are both supplied with 
seed photons by ``diskbb'' sources that have a common  $kT_{\rm in}$.
Taking a quadratic sum of  the two ``compps'' normalizations (table~2),
the  innermost disk radius of the seed photon source is 
obtained as $r^{\rm comp}_{\rm in} \sim 56 /\sqrt{\cos(i)}$ km.

The XIS spectra require an additional soft-excess component 
to be added to the power-law continuum (\S~3.4.1).
We also expressed it in terms of a simple  ``diskbb''  model,
and required it to have the same innermost temperature
as the seed photon disk.
This modeling has yielded $r_{\rm in} \sim 17^{+8}_{-17} /\sqrt{\cos(i)}$ km 
together with $kT_{\rm in} \sim$ 0.18 keV.
(Although this error region of $r_{\rm in}$ includes no-''diskbb''
assumption, the spectrum of the first half (table~3) requires significant
``diskbb'' component.)
Therefore,  based on the assumption made in \S~3.4.3,
substantial part of the cool accretion disk is thought to be covered by the Comptonizing cloud,
while the remaining  part [$\propto (r_{\rm in})^2$] is
directly visible as the soft excess.
The overall innermost radius then becomes $\sim 59/\sqrt{\cos(i)}$ km,
as a quadrature sum of $r_{\rm in}^{\rm comp}$ and $r_{\rm in}$.
Assuming an inclination angle of  $i = 70^{\circ}$ \citep{Hooft98},
the physical size  of the innermost disk radius becomes  $\sim$ 100 km,
or $\sim 6~R_{\rm s}$ for a black hole of 6$M_{\odot}$ \citep{Shahbaz99}.
The estimated disk radius is a factor of 4--7 larger 
than the values inferred from observations of this object
in the high/soft state \citep{Sala07, Kubota01}.

As argued so far, the ``diskbb'' interpretation of the soft excess
leads to a view that the optically-thick disk has retreated outwards,
by up to an order of magnitude in radius, 
compared to its location during the high/soft state.
This is consistent with the general picture regarding the relation
between these two states.
Nevertheless, the estimated inner radius of the cool disk is
small enough to suggest that a considerable fraction of
seed photons to the Compnotization process is provided by the
inner region of the cool disk which is immersed in the hot corona.
Then, the Compton cloud would be cooled by its own radiation,
which are intercepted by the cool disk, reprocessed there,
and re-emitted into the cloud. In such a case, the photon index
of the Comptonized component is expected to become $\Gamma \sim 2$
(Haardt \& Maraschi 1993), though depending on the ratio of
energy inputs to the disk and the hot cloud (Gilfanov et al. 2004).
Since we observed a much flatter continuum ($\Gamma \sim 1.4$),
some modification to the scenario might be needed (e.g., Beloborodov 1999).

Although the ``diskbb'' modeling of the soft excess gives
a natural and self-consistent interpretation,
it may be explained in alternative ways.
For example, the excess may result from scattering of 
the soft part of the continuum  by interstellar dust grains.  
Actually,  \citet{Greiner05} detected a dust scattering halo around GRO J1655$-$40,
and \citet{Sala07} reported changes in the intensity 
of the scattered component in 24 hours. 
Yet another possibility is that the soft excess is 
intrinsic to the hard continuum itself,
and produced, for example, by the ``shots'' mentioned
 in the previous section (\S~4.1).
Then, the observed change in the soft excess intensity 
could be attributed to variations in the average shot rate.

\subsection{ Iron  Line Features}

The broad emission structure at 5--8 keV has
an equivalent width of $\sim 80$~eV,
when modeled with ``diskbb+compps+compps+reflection+gau''.
The reflection component in the ``compps'' model
does not include Fe-K line emission so the broad feature in
the data is being modeled by both the complexity in the reflection component and
the fluorescent iron lines, 
presumably smeared by the disk motion.

In previous observations of GRO~J1655$-$40 made in the high/soft state,
narrow Fe-K absorption lines were detected repeatedly,
with equivalent widths of, for example, 61$^{+15}_{-13}$ eV \citep{Ueda98}, 
60--110 eV \citep{Sala07}, 
and 60--80 eV (Fe XXV+Fe XXVI; \cite{DiazTrigo07}). 
In contrast, the present Suzaku observation conducted in the low/hard state
has shown that any such feature has to have
an equivalent width  $< 25$ eV,
which is considerably smaller than values observed in the high/soft state.
The present upper limit on the Fe XXV K$\alpha$ line 
equivalent width constrains the column 
density of the corresponding ions to $< 10^{19}$ cm$^{-2}$ from a curve
of growth \citep{Ueda98},
assuming  that the absorbing plasma has a temperature of 0.1 keV.
This value is more than an order of magnitude lower
than that estimated in the high/soft state.
Further, considering that the detections of Fe-K absorption line 
features from BHBs (including GRO~J1655$-$40 itself)
have all been achieved in the high/soft state,
the present non-detection can be most naturally attributed 
to  the fact that the source was in the low/hard state.

In the high/soft state,
the Fe-K absorption features are considered to be formed by a highly ionized 
absorber, that is located at a distance 
of (5--10) $\times 10^{9}$~cm \citep{Sala07, Ueda98}, 
or $\sim 5 \times 10^{8}$ cm \citep{Miller06} from the central black hole.
It is suggested that 
such a line-forming absorber, present in the high/soft state,
undergoes significant geometrical and/or physical changes
when the system makes a transition into the low/hard state.
Then, the absorber may become 
either fully ionized from illumination by the intense X-rays,
or too tenuous (for unknown reasons),  
or too turbulent, so that the formed line features are smeared out.

\bigskip

{\it Acknowledgment} 

We thank Drs. Andrzej A.\ Zdziarski and Chris Done for useful comments 
on the  thermal-Comptonization models.
We acknowledge that a part of this work is
supported by the project of the Research Center of the Advanced
Measurement at Rikkyo University,
and the Grant-in-Aid for Specially Promoted Research (Grant No. 14079101).
S.K. gratefully acknowledges the financial support of
the Grant-in-Aid for Scientific Research (Grant No. 14654039 and
No. 15037208). 
SN acknowledges the support by JSPS (Japan Society for
the Promotion of Science) post doctoral fellowship
for foreign researchers (P05249).

\newpage

\appendix
\section*{HXD Background Subtraction with Blank-Sky Observations}

To detect signals from faint sources,
reliable background subtraction from observed data is essential.
Compared with previous instruments,
the HXD has a factor of several times lower background level
with efficient background-subtraction systems
and can achieve a higher sensitivity,
even though the background has an uncertainty of $\sim$5\% \citep{hxd1}.
Since the HXD does not have capabilities of imaging
or rocking of the detector to observe source-free regions,
we need to estimate the background using a background model 
\citep{hxd2}.
In the current analysis, we used two types of background
for subtraction.
Namely, that from a blank-sky observation and that from modeled background,
since our data analysis was begun before the modeled background was available.
Thus, we examine here the systematic error of the 
former background-subtraction method,
namely using data from
the observation of BD+30$^\circ$3639 made one day before that
of GRO~J1655$-$40,
based on the investigation of \citet{hxd2}. 
The latter subtraction method,
using a background model, has been developed by the HXD instrument team
\footnote{http://www.astro.isas.jaxa.jp/suzaku/analysis/hxd/hxdnxb/}
and the detailed properties of the modeling are described in another
paper.

As described in $\S$~3.1, the source flux of GRO~J1655$-$40 is so high
that the systematic error of the PIN background is 
considered to be negligible.
On the other hand, since the background of the GSO detector is higher
than that of the PIN,
and the flux from celestial objects becomes fainter
in the higher energy band,
the source signal of GRO~J1655$-$40 is 25\% and 10\%
of the background flux in the 40--100 keV and 100--300 keV bands, respectively.
Thus, we describe the systematic error of the GSO background
averaged over one day in the present GRO J1655$-$40 analysis,
by dividing the background into some components
and studying the variation of each component.
Note that the flux of the Cosmic X-ray background is 2-3 orders of magnitude
smaller than that of the GSO background, and it is negligible.

There are five components that affect the estimate of the GSO background 
spectrum in the GRO~J1655-40 analysis \citep{hxd2}:
(1) a component with a constant flux,
namely natural radioactive isotopes included in the detector;
(2--4) components from radio-activated sources
that are accumulated through the passages of the South Atlantic Anomaly (SAA)
and have half-lives longer than a day [denoted as ``long-nuclides'' in \citet{hxd2}], from $\sim$ 100 min to a day (``middle-nuclides''), and of $\lesssim$ 100 min (``short-nuclides''),
and (5) a component that changes in magnitude
due to the flux of cosmic-ray particles,
depending on the Cut-Off Rigidity (COR).

Although component (1) constituted $\sim$ 10\% of the 
total background flux on 2005 September 22 \citep{hxd2},
its spectrum has already been measured during ground calibration
and is temporally stable \citep{hxd1}.
Therefore, there is a very small systematic error
($\lesssim$ 0.1\%) in the total background flux
induced in the background by component (1).

In the case of component (2), the growth curves of both line 
and continuum components have been well-studied by \citet{hxd2}.
Although the line components at 70 and 150 keV are the
ones that grow most significantly,
their half-lives are $\sim$ 240 days
and the difference between adjacent days is only 
$\exp(1/240) \sim 0.3$\%/day.

The flux of component (3) changes moderately on a timescale of days,
reflecting the number of high energy particles irradiated during SAA
passages.
The number is counted by 
the PIN upper-discriminator (UD) 
counter even during the SAA \citep{hxd1}, and has a 50-day cycle that comes
from the orbital modulation between the spacecraft and the SAA \citep{hxd2}.
Figure~21 of \citet{hxd2} shows that the PIN UD amplitude of the 50-day
cycle is $\sim$ 30\% (65$\pm$10 counts s$^{-1}$), and we can estimate
that the systematic error of the background caused by component (3)
is 0.3 / (50/2 days) $\sim$ 1.2\%/day. 
The observation time of GRO~J1655$-$40 (2005 September 22) corresponds 
to a declining phase of the modulation \citep{hxd2}, 
and the background is over-subtracted by $\sim$ 1.2\% when 
that background was obtained 
from the BD+30$^\circ$3639 observation, 
which was taken one day before the GRO~J1655$-$40 observation.

\begin{figure}
  \begin{center}
\FigureFile(80mm,){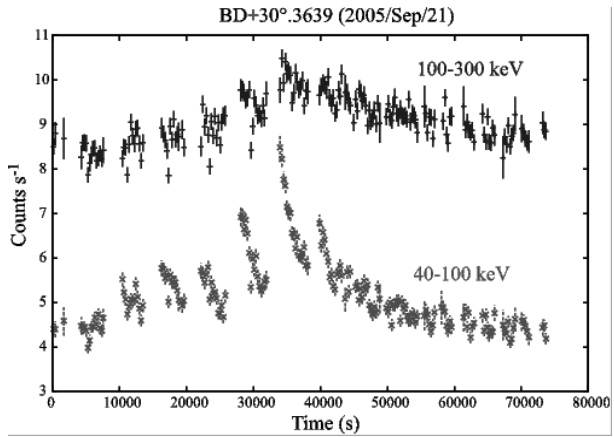}
  \end{center}
  \caption{
 Blank-sky GSO light curves in the  40--100 keV and 100--300 keV bands,
 recorded during the observation of BD+30$^\circ$3639.
 }
\label{fig:bd30_gso_lc}\label{fig:app_lc_bd30}
\end{figure}

\begin{figure}
  \begin{center}
\FigureFile(80mm,){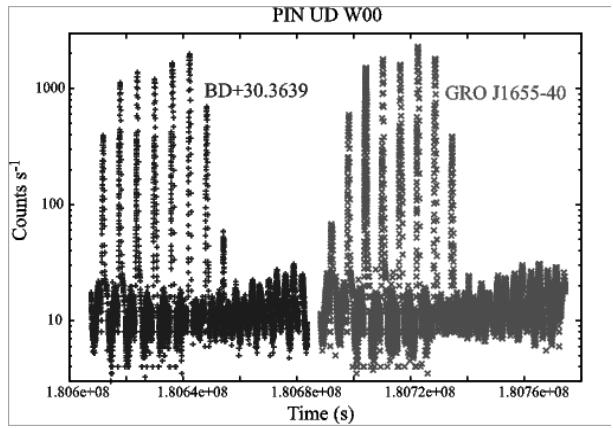}
  \end{center}
  \caption{Light curves of the PIN upper-discriminator counts 
from the HXD Unit 0, obtained from
 the observations of GRO J1655$-$40 (black) and BD+30$^\circ$3639  (grey). }
\label{fig:1655_bd30_pinud00}
\end{figure}

The light curve of the blank-sky background observation 
(BD+30$^\circ$3639) is shown in
figure~\ref{fig:app_lc_bd30} in two different energy bands, 
40--100 keV and 100--300 keV,
for which the data processing was the same as that described in $\S$~3.1.
The former data, where the flux increases periodically, 
are from orbits when the satellite went through the SAAs,
and the variation is caused by component (4).
The background variation in the lower 40--100 keV band is more significant
than that in the 100--300 keV band.  
See \citet{hxd2} for the energy spectrum of component (4).
The rate increases by at most a half of the constant flux
in the 40--100 keV band over a duration of half a day.
Assuming that the shape of the increase
in count rate of the background light curve is a triangle over half a day,
the additional amounts of the background caused by (4) are obtained 
as $0.5/2/2 \sim 13\%$ of the constant flux over a one-day period.
The flux of component (4) also depends on the number of 
particles irradiated during each SAA,
and the difference between the GRO~J1655$-$40 and 
BD+30$^\circ$3639 observations is directly comparable
with the PIN-UD counts measured in the two observations, 
since the half-life of the component (4) is smaller than $\sim$ 100 min.
Figure~\ref{fig:1655_bd30_pinud00} 
shows the PIN UD rates of the WELL Unit 00 \citep{hxd1}
during the observations of GRO J1655$-$40 and BD+30$^\circ$3639. 
The observed total PIN UD counts are $7.30 \times 10^{7}$ for GRO J1655$-$40
and $7.70 \times 10^{7}$ for BD+30$^\circ$3639.
The difference is 5\%,
and the integrated start and stop times of the two datasets 
were adjusted to overlap each other.
Therefore, we can estimate that the systematic error caused by
component (4) is
$0.13 \times 0.05 \sim 0.7\%$ of the total background count rate
in the 40--100 keV band.
In the higher 100--300 keV band,
the variation amplitude of component (4) is
about half of that in the 40--100 keV band,
resulting in a systematic error of $\sim$ 0.4\% in the total background flux.
Since the distribution of the time after the SAA passage was adjusted
between the two observations as described in $\S$~3.1,
it does not affect the systematic error of the background.

\begin{figure}
  \begin{center}
\FigureFile(80mm,){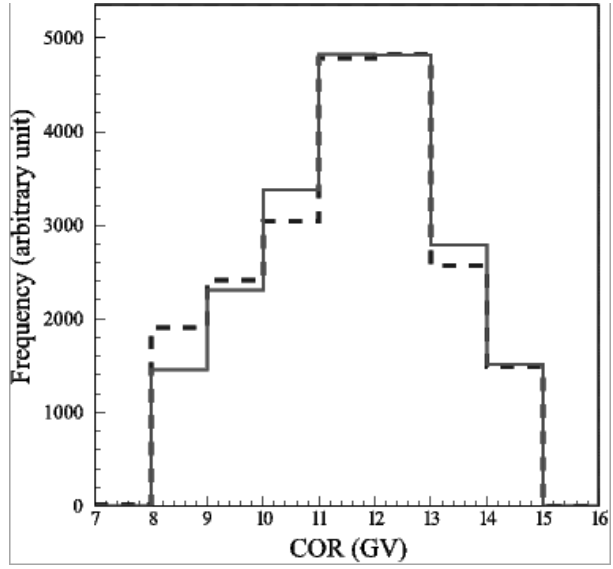}
  \end{center}
  \caption{
  Distributions of COR during the observations of 
  GRO J1655$-$40 (solid) and BD+30$^\circ$3639 (broken). 
 The periods during which the sources were occulted by the Earth 
were discarded. }
\label{fig:1655_bd30_cor}\label{fig:app_dist_cor}
\end{figure}

The background variation in the latter half of figure~\ref{fig:app_lc_bd30} 
is caused by a difference in the COR \citep{hxd2}, 
and it is considered that component (5) constitutes $\sim$~10\%  
and $\sim$~5\% of the total flux in the background in the 40--100 keV 
and 100--300 keV energy band respectively.
See \citet{hxd2} for details of the COR dependence.
Figure~\ref{fig:app_dist_cor} compares the COR distributions
between the GRO J1655$-$40 and BD+30$^\circ$3639 observations. 
The distributions for the two cases are very similar,
with average values of 11.9 and 11.7 GV, respectively.
The total of the difference of the exposure time in every COR bin is 
$\sim$ 600 s,
which is $\sim$ 3\% of the total observation time ($\sim$ 20 ks).
Even if the differences in the background flux for these
terms are all 10\% of the total background flux in the 40--100 keV band, 
the systematic error induced by component (5) is
considered to be $0.1 \times 0.03 \sim 0.3\%$ of the total background 
in the 40--100 keV band. 
Similarly, the error is $0.05 \times 0.03 \sim 0.2\%$ in the 100--300 keV
band.

From the above discussion, the systematic errors in the 
GRO~J1655$-$40 background obtained from the BD+30$^\circ$3639 dataset 
are summarized for each component
as (1) $\lesssim$ 0.1\%, (2) +0.3\%, (3) -1.2\%, 
(4) -0.7\% over 40--100 keV and -0.4\% over 100--300 keV, 
and (5) at most -0.3\% over 40--100 keV and -0.2\% over 100--300 keV,
where the symbol "+" represents an under-subtraction and 
the "-" symbol indicates over-subtraction.
Therefore, adding up all the components, the total systematic error 
of the current background is thought to be at 
most -1.9\% and -1.5\% in the 40--100 keV and 100--300 keV bands respectively.

In order to confirm this estimation, 
we additionally examined three sets of GSO backgrounds
obtained during two successive days near the GRO~J1655$-$40 observation.
Namely,
a BD+30$^\circ$3639 observation on 2005 September 21 and
a HESS~J1616-508 observation on 2005 September 20,
an RX~J1713-39 observation on 2005 September 26 and 25, 
and a continuous observation of the same source on 2005 September 28 and 27.
All the datasets were obtained in the same declining phase of the
50-day cycle of the PIN-UD modulation.
We extracted the spectra by the same method as in $\S$~3.1, and we 
also subtracted the background of the previous day from that of the next day.
The result shows that the differences between the two observations are
all within -1.9~$\sim$~+1.5\% and -0.9~$\sim$~0.0\% of the count rate 
in the 40--100 keV and 100--300 keV bands respectively.

Following the estimation and the result, we assume the peak-to-peak 
systematic error of the one-day average GSO background spectrum 
obtained from the adjacent dataset is $\pm$2\% 
in the present GRO~J1655$-$40 analysis, 
although the error in the 100--300 keV band could be smaller by about a half.
This 2\% background uncertainty results in 8\% and 20\% systematic errors 
in the GRO~J1655$-$40 source spectrum in the 40--100 keV and 100--300 keV
bands respectively, since the source signal is 25\% and 10\% 
respectively of the background flux in these energy bands.

\clearpage

\begin{table*}
\begin{center}
\caption{
Best fit parameters  obtained with the average spectra from XIS and HXD
 data, independently. $^{\ast,\dag,\ddag}$
}
\begin{tabular}{lccccc}
\hline
\hline
\multicolumn{6}{l}{Data, Model} \\
\hline
         & pow / cutoffpl      & diskbb                        &gau       & wabs             & $\chi$$^2$/d.o.f\\
         & $L_{\rm x}$$^{\S,\P}$       & $L_{\rm x}$$^{\S,\P}$ & $E_c$ (keV)        & $N_{\rm H}$$^{\L}$ & \\
         & ${\it \Gamma}$   & $r_{\rm in}$ (km)$^{\P,\#}$  & $\sigma$ (keV)  &         & \\
         &  $kT_{\rm cut}$  & $kT_{\rm in}$ (keV)   & $E.W.$ (eV) &     & \\
\hline
\multicolumn{6}{l}{All data (XIS), diskbb+pow+gau} \\
\hline
         & 3.70                             & 0.07            & 6.50$\pm$0.08           & 7.4$^{+0.2}_{-0.1}$ & 1499/1419 \\
        & 1.75$\pm$0.01                   & 45$^{+9}_{-8}$   & 1.04$^{+0.13}_{-0.10}$              &                    & \\
        & $\cdots$                        & 0.18$\pm$0.01   & 280$\pm$40           &                    & \\
\hline
\multicolumn{5}{l}{All data (PIN+GSO), cutoffpl}\\
\hline
        & 4.61                             &   $\cdots$      &  $\cdots$            & $\cdots$          & 64/67\\
        &  1.35$\pm$0.04                   & $\cdots$        &  $\cdots$            &                   &      \\
        &  200$^{+50}_{-30}$                & $\cdots$      &  $\cdots$             &                   &      \\
\hline
\multicolumn{5}{@{}l@{}}{\hbox to 0pt{\parbox{180mm}{\footnotesize
$^{\ast}$ Errors refer to 90\% confidence limits.\\
$^{\dag}$ The energy offsets of the XIS spectra are adjusted within $\pm$10 eV.\\
$^{\ddag}$ The PIN and GSO background spectra are extracted from the one-day before observation (BD+30$^\circ$3639).\\
$^{\S}$ The luminosities are in the 0.7--300 keV band in the unit of 10$^{36}$ erg s$^{-1}$.\\
$^{\P}$ The distance of 3.2 kpc is assumed.\\
$^{\L}$ In the unit of 10$^{21}$ H atoms cm$^{-2}$.\\
$^{\#}$ The value should be divided by  ${\sqrt{\cos(i)}}$,
where $i$ is the inclination of the disk.\\
}\hss}}
\end{tabular}
\end{center}
\end{table*}

\begin{table*}
\begin{center}
\caption{
Best fit parameters obtained with the 0.7--300 keV average spectra (XIS+PIN+GSO) $^{\ast,\dag,\ddag}$
}
\begin{tabular}{lccccccc}
\hline
\hline
\multicolumn{8}{l}{Data, Model} \\
\hline
         & compps (soft)                             & compps (hard)
 & diskbb                               &  reflection $^{\S}$
 & gau               & wabs             & $\chi$$^2$/d.o.f\\
         & $L_{\rm x}$$^{\P,\L}$                       & $L_{\rm
 x}$$^{\P,\L}$                        & $L_{\rm x}$$^{\P,\L}$   &
 $L_{\rm x}$$^{\P,\L}$     & $E_c$ (keV)   & $N_{\rm H}$$^{\#}$ & \\
         & $y$, $r_{\rm in}^{\rm comp}$ (km)$^{\L,\dag\dag}$   & $y$,
 $r_{\rm in}^{\rm comp}$ (km)$^{\L,\dag\dag}$   & $r_{\rm in}$ (km)$^{\L,\dag\dag}$    
 & $\Omega/2\pi$, $R_{\rm in}$ ($R_{\rm s}$)     & $\sigma$ (keV) &
 & \\
         & $kT_{\rm e}$ (keV)                         & $\cdots$
 $^{\ddag\ddag}$                      & $kT_{\rm in}$ (keV)         &
 $\xi$ (erg cm s$^{-1}$)                      &  $E.W.$(eV)                 & \\
\hline
\multicolumn{7}{l}{All data, diskbb+compps+gau$^{\S\S}$} \\
\hline
        &  3.86                         & $\cdots$
 & 0                             & $\cdots$
 & 6.4 & 5.4                     & 2408/1489 \\
        &  0.9, 94                      & $\cdots$, $\cdots$                        & 0                           & $\cdots$, $\cdots$                         &    0.5 &                &            \\
        &  60                           & $\cdots$                                  & 0.10                              & $\cdots$                                   &        70 &            &            \\
\hline
\multicolumn{8}{l}{All data, diskbb+compps+reflection+gau$^{\S\S}$} \\
\hline
       & 3.56                           & $\cdots$                                  & 0                              & 0.44                                & 6.4 &  5.5       & 2246/1486 \\
       & 0.9, 94                        & $\cdots$,  $\cdots$                       & 0                           & 0.4, 6                             &   0.5 &                 & \\
       & 70                             & $\cdots$                                  & 0.10                              & 0                                     &   65 &                 & \\
\hline
\multicolumn{8}{l}{All data, diskbb+compps+compps+gau} \\
\hline
       & 0.82                          & 4.26                                      & 0.06                             & $\cdots$                             &  6.4$^{+0.1}_{-0.2}$ &7.2$^{+0.2}_{-0.3}$ & 1600/1487 \\
       & 0.5$\pm$0.1, 34$\pm$4         & 1.3$\pm$0.1, 18$\pm$2                          &  40$^{+4}_{-6}$   & $\cdots$, $\cdots$                    &   0.7$\pm$0.1 &                 & \\
       & 55$\pm$5                      & $\cdots$                                  & 0.18$\pm$0.01    & $\cdots$                              &  110 &                  & \\
\hline
\multicolumn{7}{l}{All data, diskbb+compps+compps+reflection+gau} \\
\hline
       & 0.79                         & 3.63
 & 0.01                          & 0.07(soft), 0.63(hard)                 & 6.2$^{+0.2}_{-0.1}$ & 7.4$^{+0.1}_{-0.2}$ & 1524/1484 \\
       & 0.28$^{+0.11}_{-0.05}$, 51$^{+7}_{-5}$   & 1.3$^{+0.2}_{-0.1}$,
 23$\pm 3$                 &   17$^{+8}_{-17}$                 & 0.6$\pm 0.1$, $>600$      & 0.6$\pm$0.1 &                   & \\
       & 140$^{+5}_{-10}$              & $\cdots$
 & 0.18$\pm$0.01                       & 350$\pm$100                              &   80 &                 & \\
\hline
\multicolumn{7}{@{}l@{}}{\hbox to 0pt{\parbox{180mm}{\footnotesize
$^{\ast}$ Errors refer to 90\% confidence limits.\\
$^{\dag}$ The energy offsets of the XIS spectra are adjusted within $\pm$10 eV.\\
$^{\ddag}$ The PIN and GSO background spectra are extracted from the one-day before observation (BD+30$^\circ$3639).\\
$^{\S}$ The reflection of the both ``compps'' components with the same
 parameters for two ``compps'' model.\\
$^{\P}$ The luminosities are in the 0.7--300 keV band in units of 10$^{36}$ erg s$^{-1}$.\\
$^{\L}$ A distance of 3.2 kpc is assumed.\\
$^{\#}$ In units of 10$^{21}$ H atoms cm$^{-2}$.\\
$^{\dag\dag}$ The values should be divided by  ${\sqrt{\cos(i)}}$,
where $i$ is the inclination of the disk.\\
$^{\ddag\ddag}$ The $kT_{\rm e}$ values are the same between the two ``compps'' components.\\
$^{\S\S}$ Errors are not shown because of large $\chi^{2}$ values.\\
}\hss}}
\end{tabular}
\end{center}
\end{table*}


\begin{table*}
\begin{center}
\caption{
The same as table~2 but using the first-half and second-half spectra (XIS+PIN) $^{\ast,\dag,\ddag}$
}
\begin{tabular}{lccccccc}
\hline
\hline
\multicolumn{7}{l}{Data, Model} \\
\hline
         & compps (soft)                             & compps (hard)
 & diskbb                               &  reflection $^{\S}$         & gau                       & wabs             & $\chi$$^2$/d.o.f\\
         & $L_{\rm x}$$^{\P,\L}$                       & $L_{\rm
 x}$$^{\P,\L}$                        & $L_{\rm x}$$^{\P,\L}$   &
 $L_{\rm x}$$^{\P,\L}$  & $E_c$~(keV)      & $N_{\rm H}$$^{\#}$ & \\
         & $y$, $r_{\rm in}^{\rm comp}$ (km)$^{\L,\dag\dag}$   & $y$, $r_{\rm in}^{\rm comp}$ (km)$^{\L,\dag\dag}$   &     $r_{\rm in}$ (km)$^{\L,\dag\dag}$   & $\Omega/2\pi$, $R_{\rm in}$ ($R_{\rm s}$)     &     $\sigma$~(keV))      &       & \\
         & $kT_{\rm e}$ (keV)                         & $\cdots$
 $^{\ddag\ddag}$                      &$kT_{\rm in}$ (keV)            &
 $\xi$ (erg cm s$^{-1}$)         & $E.W.$(eV)             &                   & \\
\hline
\multicolumn{7}{l}{First half, diskbb+compps+compps+reflection+gau} \\
\hline
       & 0.90                      & 3.80                              &
 0.02                          & 0.10(soft)+0.68(hard)     & 6.3$^{+0.2}_{-0.1}$            & 7.4$\pm$0.1 & 808/798 \\
       & 0.28$^{+0.09}_{-0.08}$, 55$^{+7}_{-6}$  & 1.3$^{+0.3}_{-0.1}$,
 23$^{+2}_{-3}$   & 22$\pm$8                    & 0.6$^{\S\S}$,
 900$^{\S\S}$       &  0.7$\pm$0.1 &                 & \\
       & 135$^{+5}_{-10}$           & $\cdots$
 &  0.18$\pm$0.01                    & 350$^{\S\S}$             & 100                 &                   & \\
\hline
\multicolumn{7}{l}{Second half, diskbb+compps+compps+reflection+gau} \\
\hline
       & 1.00                     & 3.27                          &
 $<0.02$                          & 0.12(soft)+0.60(hard)  & 6.2 $\pm$0.1               & 7.4$^{+0.2}_{-0.1}$ & 880/924 \\
       & 0.31$^{+0.11}_{-0.05}$, 52$^{+7}_{-5}$   & 1.4$^{+0.4}_{-0.1}$,
 28$^{+2}_{-3}$   &  $<$21   & 0.6$^{\S\S}$, 900$^{\S\S}$       &
 0.6$\pm$0.1 &                & \\
       & 155$\pm 15$          & $\cdots$                      &
 0.18$\pm$0.01                      & 350$^{\S\S}$            & 90                  &                   & \\
\hline
\multicolumn{7}{@{}l@{}}{\hbox to 0pt{\parbox{180mm}{\footnotesize
$^{\ast}$ Errors refer to 90\% confidence limits.\\
$^{\dag}$ The energy offsets of the XIS spectra are adjusted within $\pm$10 eV.\\
$^{\ddag}$ The PIN background spectra are extracted from the one-day before observation (BD+30$^\circ$3639).\\
$^{\S}$ The reflection of the both ``compps'' components with the same
 parameters for two ``compps'' model.\\
$^{\P}$ The luminosities are in the 0.7--300 keV band in units of 10$^{36}$ erg s$^{-1}$.\\
$^{\L}$ A distance of 3.2 kpc is assumed.\\
$^{\#}$ In units of 10$^{21}$ H atoms cm$^{-2}$.\\
$^{\dag\dag}$  The values should be divided by  ${\sqrt{\cos(i)}}$,
where $i$ is the inclination of the disk.\\
$^{\ddag\ddag}$ The $kT_{\rm e}$ values are the same between the two ``compps'' components.\\
$^{\S\S}$ Reflection parameters are fixed to those of the average spectrum in table~1.\\
}\hss}}
\end{tabular}
\end{center}
\end{table*}


\clearpage

\newpage


\end{document}